\documentclass[vecphys]{svmult}

\usepackage{makeidx}         
\usepackage{graphicx}        
\usepackage{multicol}        
\usepackage{cite}            
\usepackage[bottom]{footmisc}
\usepackage{amsmath}

\makeindex             
\usepackage{color}
\usepackage{siunitx}
\newcommand{\todo}[1]{\marginpar{\small To do: \color{red}#1}}
\newcommand{\ket}[1]{\ensuremath{\left|\mathrm{#1}\right\rangle}}
\newcommand{\bra}[1]{\ensuremath{\left\langle\mathrm{#1}\right|}}

\begin{document}

\title{Time-bin Entanglement from Quantum Dots}
\author{Gregor Weihs$^1$ \and Tobias Huber$^1$ \and Ana Predojevi\'c$^2$}
\institute{Universit\"at Innbruck, Institut f\"ur Experimentalphysik, Technikerstr. 25, 6020 Innsbruck, Austria
 \texttt{gregor.weihs@uibk.ac.at}
 \and Institute for Quantum Optics, University Ulm, Albert-Einstein-Allee 11, 89081 Ulm, Germany
 \texttt{ana.predojevic@uni-ulm.de}}

\maketitle

\begin{abstract}
The desire to have a source of single entangled photon pairs can be satisfied using single quantum dots as emitters. However, we are not bound to pursue only polarization entanglement, but can also exploit other degrees of freedom. In this chapter we focus on the time degree of freedom, to achieve so-called time-bin entanglement. This requires that we prepare the quantum dot coherently into the biexciton state and also build special interferometers for analysis. Finally this technique can be extended to achieve time-bin and polarization hyper-entanglement from a suitable quantum dot.
\end{abstract}

\section{Introduction}

While the realization of a general purpose, universal quantum computer of a useful size appears to be some time away, small-scale and special purpose quantum computing devices have been realized or are under construction. Quantum cryptographic protocols, in particular quantum key distribution (QKD), which lets us distribute a secure cryptographic key between two parties, are already commercial to some degree. Yet the distance over which the key exchange can be realized is limited to a few hundred kilometers of optical fiber, due to the inevitable exponentially growing losses and the noise floor or background level of any realistic detector. The practical limits are even shorter because for long distances the key rates will be extremely low.

While for QKD one may resort to classical, trusted repeaters, thus sacrificing the absolute physical security of the key exchange, for connecting quantum information processing devices we will have to implement so-called quantum repeaters \cite{Briegel98a}. Quantum repeaters break a long distance connection into smaller links over which entanglement is established. Via Bell-state measurements (BSM) at the intermediate nodes the entanglement over the smaller links is then converted into entanglement between the endpoints. What sounds simple in this very abstract description is much more difficult in practice, because we must not assume perfect quantum channels even for the smaller links. While protocols \cite{Briegel98a} have been devised to cope with the errors, the resulting overhead in resources appears to be forbidding. Only if we start with a high degree of entanglement in the small links will it be feasible to establish the end-to-end quantum channel.

The most frequently used source of entanglement is the spontaneous parametric down-conversion (SPDC) source \cite{Kwiat95b,Predojevic12a}, which produces pairs of entangled photons through a nonlinear optical effect from a pump laser. Unfortunately, SPDC does not create a single entangled pair at a time, but is rather probabilistic, so that a (very) small fraction of the pump laser's photons are converted resulting in a random number of pairs per output pulse or time window. This limits their applicability in quantum repeaters, because there is a fundamental trade-off between a high pair emission rate and the error rate that is caused by multi-pair emissions. This error rate dramatically reduces the achievable distance in a multi-link repeater scenario, even at a two-pair emission probability of only 1\% \cite{Guha15a}.

This is the ultimate reason why quantum communication will eventually need sources of single entangled photon pairs. For now these are limited to single quantum emitters with cascaded optical transitions. Atoms can serve as entangled photon pair sources \cite{Freedman72a,Aspect81a,Weber09a}, but they require complex atomic beam or trap setups and their overall emission rate is limited. To our knowledge no entangled photon pairs have been produced from single molecules or color centers in solids, which otherwise seem to work well as single photon sources. This leaves semiconductor quantum dots as the only viable solid-state single quantum emitter of entangled photon pairs.

Proposed initially in Ref.~\cite{Benson00a} the biexciton-exciton cascade may emit polarization entangled photon pairs, if the two spin configurations of the intermediate exciton state are degenerate and thus no which-path information is available. The status of polarization entanglement from quantum dots is discussed in detail in chapter \textbf{XX}\todo{Insert proper cross-reference}. In this chapter we would like to point out that once we have an emitter of photon pairs, there may be other degrees of freedom available to us for realizing photon entanglement. Further we will discuss our and others' results on time-bin entanglement from quantum dots with an outlook on improvements and the possibility of generating hyperentanglement of photons in two degrees of freedom.

The chapter will start by discussing the degrees of freedom of a photon and their measurement, followed by a more detailed discussion of the related phenomena of energy-time and time-bin entanglement. We will show that a coherent excitation mechanism  is required for obtaining time-bin entanglement from a quantum dot and will discuss the optimal conditions. Finally we will present the results on time-bin entanglement and an outlook.

\section{Photon degrees of freedom}

Without resorting to a particular interpretation we may define a photon to be an elementary excitation of a quantized mode of the electromagnetic radiation field. A mode is a solution to the wave equation under particular boundary conditions and in particular we can always resort to monochromatic solutions so that the modes are harmonic solutions with a particular frequency $\omega$ and the spatial part is the corresponding solution of the Helmholtz equation. In a box-like quantization volume with fixed or periodic boundary conditions these will be plane waves. From these plane, monochromatic waves we may build other monochromatic modes by unitary transformations. In experiments these unitary transformations between different sets of modes or wavevectors are effected by beam-splitters or other, similar couplers. We may further resort to non-monochromatic or spatio-temporal modes, like wavepackets, which obviously will not necessarily be orthogonal, but in most practical cases may be constructed close to orthogonal \cite{Vogel06a}.

A plane wave is characterized by a wavevector $\mathbf k$ with magnitude $k=\omega n(\omega)/c$ and a polarization unit vector $\mathbf e$, which is orthogonal to $\mathbf k$. Because there are always two orthogonally polarized modes for any $\mathbf k$, a photon may have any state of polarization that can be described as a superposition of the two, i.e. any state on the Poincar\'e sphere.

A single photon with a given frequency and wavevector is thus a perfect two-state system, or qubit, with degenerate energy levels. The polarization of a photon can be manipulated easily using retarders (wave-plates) and measured using polarizers, which effectively project any incoming polarization to the one transmitted. Polarizing beam-splitters (PBS), also called two-channel polarizers, are devices that couple polarization and spatial mode (wavevector) by transmitting light that is polarized parallel to the plane of incidence (p) and reflecting light that is polarized perpendicular to the same (s).

In contrast to the polarization, which is discrete, the continuous degrees of freedom frequency and wavevector allow storing more information in one photon \cite{Tittel01a}. In most practical cases we will strive to define a discrete but not necessarily binary set of modes for transmitting and manipulating photonic quantum information, because the analysis in the presence of noise and distortion through a channel will become difficult for continuous encoding. Several schemes have been put forward and demonstrated for wavevector-spatial coding: the dual-rail qubit \cite{Ralph10a} and its multi-rail extension, transverse paraxial mode coding, in particular orbital angular momentum (OAM) \cite{Yao11a} and similar rotationally invariant coding \cite{DAmbrosio12a}. In the frequency/energy-time dimension time-bin \cite{Marcikic02a} and multi-time-bin coding have been used as well as generalized temporal mode \cite{Xing14a,Brecht15a} and frequency mode coding \cite{Zavatta14a}.

Some of these encodings promise good stability of the quantum state under propagation either in free space or in optical waveguides. On the other hand their manipulation and analysis (decoding) present more difficulties than in the simple case of polarization. In all cases one has to deal with some kind of interferometer. As an example, for OAM decoding only recently efficient methods were found \cite{Mirhosseini13a}. In practice one will thus choose an encoding that is robust for the chosen channel. There seems to be some general agreement that the most robust encoding for  long-distance transmission of quantum information in optical fibers is time-bin encoding or some variant of it, e.g. differential phase shift keying (DPSK).

\section{Time-bin encoding and entanglement}
\label{sec:time-bin}

In classical communication a large variety of modulation schemes is known, both in incoherent and coherent communication, analog and digital. For photons, any classical scheme can be used or adapted in principle. The only thing that changes are the fundamental noise limits given by the uncertainty principle for amplitude and phase.

\begin{figure}
  \centering
  \includegraphics[width=\textwidth]{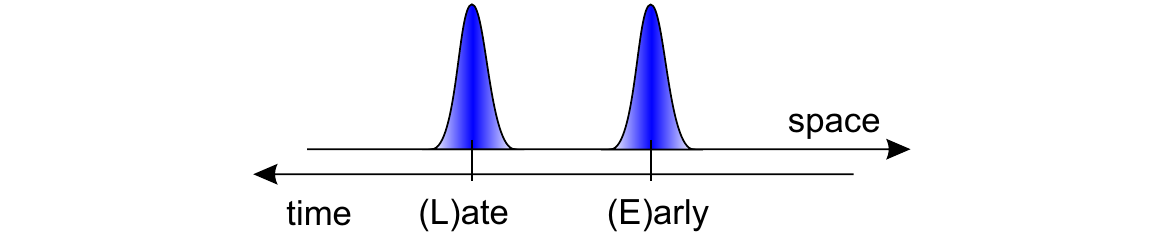}
  \caption{Time bins are quasi-orthogonal wavepacket envelopes, which form the two quantum states of a photonic qubit when occupied by a single photon. In this picture we imagine the wavepackets propagating to the right in real space. The photon can then be in either the early (E) or late (L) state or any superposition thereof.} \label{fig:pulses}
\end{figure}

In this sense time-energy wavefunctions and time-bin qubits are particular quantum variants of classical phase-shift-keying (PSK), even though in the quantum realm we rarely use continuous-wave carriers. A time-bin qubit is defined via two usually pulse-like quasi-orthogonal temporal wavepackets as shown in Fig.~\ref{fig:pulses}. A general pure state is thus $\ket\psi=\alpha \ket E + \beta \ket L$. The superposition bases $\ket E \pm \ket L$ are sometimes called energy bases, even though this terminology is only accurate if we are talking about energy (frequency) eigenstates, i.e. plane waves, which are complementary to a time basis with temporal $\delta$-distributed wavefunctions. The particular wavepacket shape will either be determined by the generating optical (laser) pulse or the decay properties of the generating quantum emitter.

Before discussing advantages and disadvantages of the time-bin encoding we would like to look at the historical perspective. Temporal superpositions of photons were first proposed by Franson in Ref.~\cite{Franson89a} in the context of entanglement and Bell's inequalities. To our knowledge, his original proposal of using cascaded transitions in atoms was never realized to generate energy-time entangled photon pairs or demonstrate a violation of Bell's inequality. Most experiments \cite{Brendel92a,Kwiat93a,Brendel99a} used SPDC as the photon pair source in which the coherence of the pump laser provides the coherent superposition of the early and late times. The requirement for using discrete time bins through pulsed pumping of the SPDC instead of the continuous variant derived from the desire to use the so-encoded qubits in protocols that require interferometric Bell-state analysis on photons from different sources such as quantum teleportation \cite{Bouwmeester97a} and entanglement swapping \cite{Jennewein02a}.

\begin{figure}
  \centering
  \includegraphics[width=\textwidth]{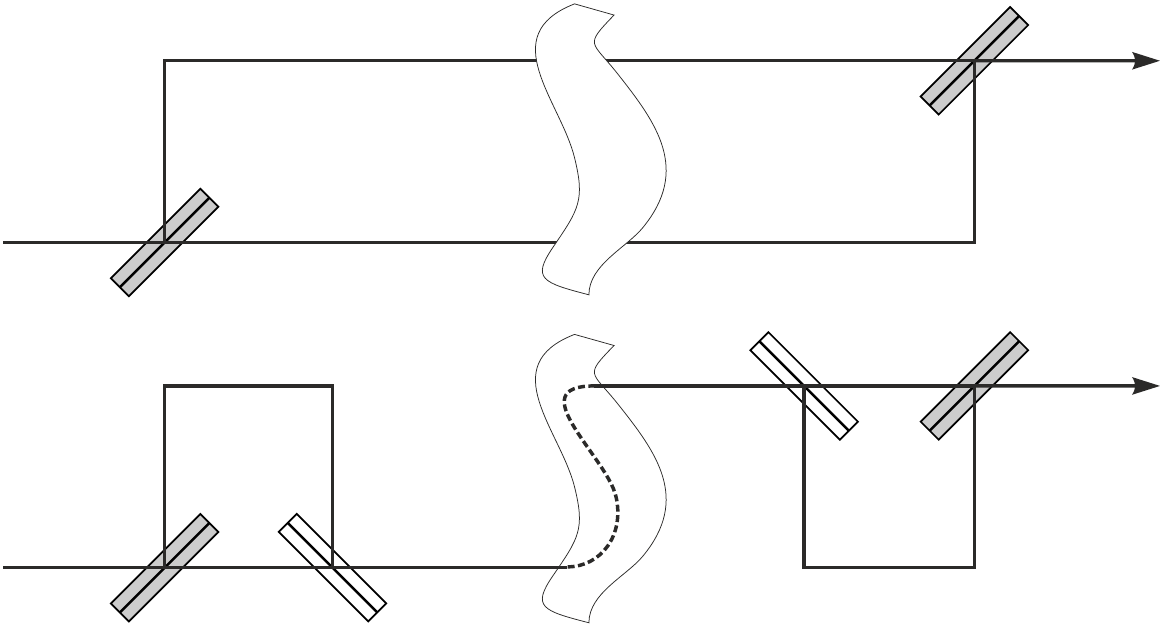}
  \caption{Top: the first beamsplitter creates a superposition of a photon being either in the upper mode (rail) or the lower one. The second one analyzes the superposition depending on the phase difference accrued between the two paths along their entire length, or in other words, it converts the superposition back to the photon going to either of its outputs. Bottom: additional mirrors (not shown) and beamsplitters multiplex the upper mode onto the the lower one with some delay. The second imbalanced interferometer undoes the delay for measurement. This works only probabilistically, with 50\% efficiency in each interferometer, i.e. 25\% overall for analysis in superposition bases. Better performance could be achieved by using a switch.}
  \label{fig:multiplex}
\end{figure}

For long distance quantum communication through optical fibers time-bin encoding has a decisive advantage. The relative phase between two pulses that are only a few nanoseconds apart in time is only altered by changes in the environment that are faster than their temporal separation, i.e. in the GHz range. This kind of encoding can also be seen as a temporally multiplexed version of the dual-rail qubit (see Fig.~\ref{fig:multiplex}). Nevertheless, chromatic dispersion can play a role both through the induced pulse spreading and inside the imbalanced interferometers that are required for time-bin analysis.

\begin{figure}
  \centering
  \includegraphics[width=\textwidth]{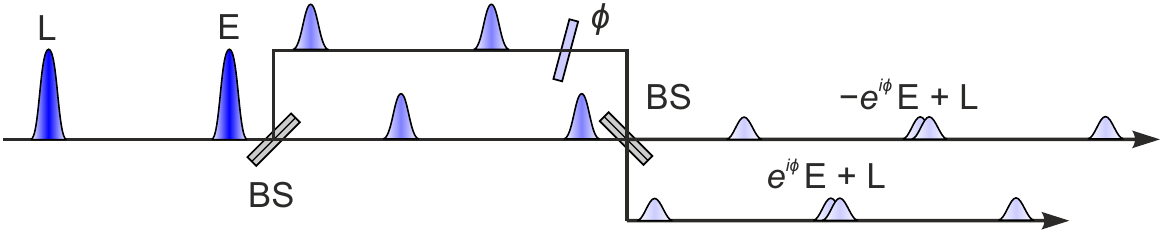}
  \caption{Time-bin analysis occurs through imbalanced interferometers built from two beamsplitters (BS) and mirrors (not shown). The output probability of a photon is distributed into three pulses, where the middle ones are the two complementary superpositions. However, only half of the total probability (photons) will be in the middle pulses. Depending on the phase of the initial state and the interferometer phase $\phi$ this probability will be distributed between the two interferometer outputs. The phase shift allows us to analyze with regard to a particular basis, i.e. $X$, $Y$ or any other equal-amplitude superposition of the E and L states. The first and third pulse contain the other half of the early and late pulses, i.e. when detected, they give projections to the E/L ($Z$) basis.}
  \label{fig:analyzer}
\end{figure}

In order to measure a time-bin qubit in a superposition basis we have to delay the early time bin and interfere it with the late one. For this purpose we use an imbalanced Mach-Zehnder or Michelson interferometer (see Fig.~\ref{fig:multiplex}). Obviously this analysis is lossy, because only 50\% of the photons will experience the correct delay. Half of the early time bin will not be delayed and half of the late time bin will be delayed even further. This results in a temporal pulse pattern as shown in Fig.~\ref{fig:analyzer}. If the beam splitters are symmetric, then only bases on the equator of the Bloch sphere (with E and L at the poles) can be analyzed. For $\phi=0$ the interferometer outputs correspond to the qubit states $\pm X$, for $\phi=i$ to $\pm Y$. More general, universal time-bin analyzers require beamsplitters with adjustable splitting ratios \cite{Bussieres10a} to allow arbitrary amplitude superpositions.

The ultimate time-bin analyzer (and encoder) uses a switch (switchable mirror) instead of the first beamsplitter. The switch would have to route the early time bin along the long arm and the late time bin along the short arm. The splitting ratio of the second beamsplitter then defines the amplitude ratio of the superposition states that are to be analyzed.

While time-bin encoding is very stable under propagation through optical communication channels, the stability of an imbalanced interferometer may be a concern. They have been realized in free-space and fiber versions and in both cases one needs to add a phase stabilization laser and ensure the best possible mechanical and thermal stability. Because of the large imbalance the stabilization laser not only needs to have a long enough coherence length but is also required to be locked in its absolute wavelength, which is almost always chosen different from the wavelength of the single photons to avoid stray laser light reaching the sensitive single photon detectors.

\begin{figure}
  \centering
  \includegraphics[width=\textwidth]{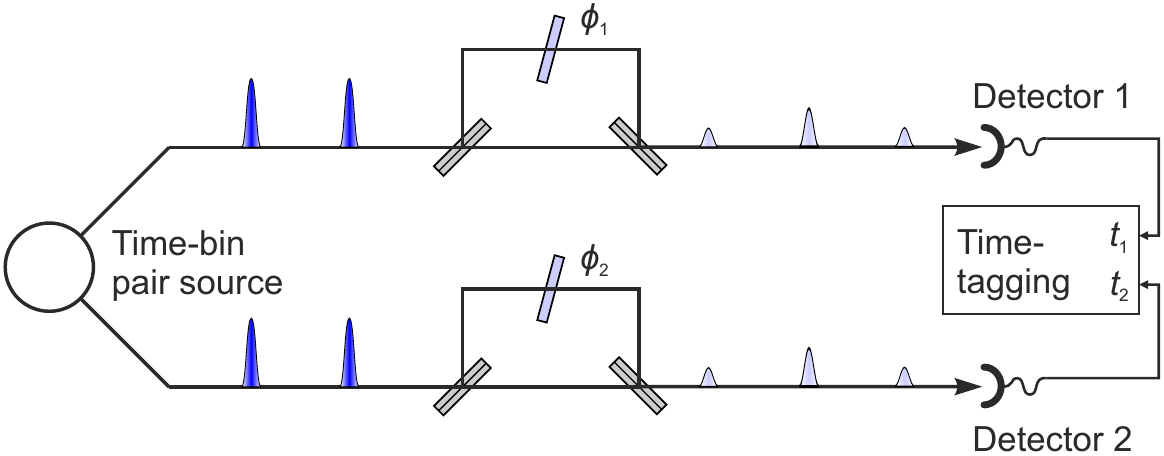}
  \caption{Analyzing time-bin encoded photon pairs requires time-bin analyzers with phase settings $\phi_1$ and $\phi_2$ to set the basis on both sides and time-correlated detection. The amount of imbalance is kept at the minimum allowed by the time resolution of the detectors or the minimum possible wavepacket duration for the source.}
  \label{fig:timebinent}
\end{figure}

So far the discussion concerned a single time-bin qubit. Things get somewhat more complicated for two qubits, which may be entangled or not. The situation is visualized in Fig.~\ref{fig:timebinent}, where a source emits pairs of time-bin encoded photons. Both photons are analyzed in identical interferometers and detectors, whose detection times are recorded as $t_1$ and $t_2$. A simple start-stop measurement between the two sides is not sufficient, as it would lump the superposition basis events in with other simultaneous detection events. Instead one needs to either record (time-tag) the arriving photons in absolute time or at least determine the time difference to a synchronization signal from the source on each side.

A perfect source of time-bin entanglement would produce the maximally entangled state
\begin{equation}\label{eq:tbstate}
  \ket{\Phi}(\phi_l)=\frac{1}{2}\left(\ket{E_1E_2}+e^{i\phi_l}\ket{L_1L_2}\right),
\end{equation}
where the phase $\phi_l$ is internal to the source. It may originate, for example, from the superposition of pump pulses in SPDC. By setting $\phi_l=0,\pi$ one thus obtains the $\mathrm\Phi^{+,-}$ Bell states, respectively. For the $\mathrm\Phi^+$ state the coincidence count rate for the two middle pulses in a pairing of two equivalent outputs of the analyzers will then vary as $\frac{1}{16}(1+\cos(\phi_1+\phi_2))$, i.e. a coincidence probability of $1/8$ for $\phi_1+\phi_2=0$. The same is true for the second equivalent pairing and both are complementary to the coincidence count rates for the inequivalent output pairings. Therefore, in total only for one quarter of all emitted pairs both photons are detected in the superposition basis, for another quarter both photons are detected in the time-bin basis and for the remaining half, one photon each is detected in the superposition and time-bin bases, respectively.

\section{Time-bin entanglement from single quantum emitters}

In the original proposal by Franson \cite{Franson89a} a long-lived upper level in an atom provided the required coherence between the early and late cascade emission of a photon pair. In SPDC the coherence of an earlier or later produced photon pair is provided by the coherence time of the pump laser, whose phase will be the sum phase of the paired photons. This can either happen with a continuous-wave laser for Franson-type entanglement or with a coherent superposition of an early and late pump pulse for time-bin entanglement. The laser pulses can be produced by an imbalanced interferometer or directly from a mode-locked laser.

\begin{figure}
  \centering
  \includegraphics[width=\textwidth]{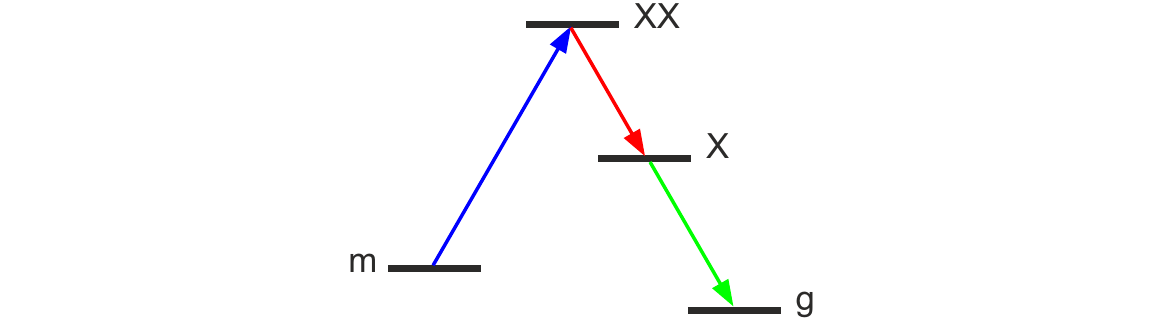}
  \caption{Bringing the quantum dot to a metastable level before exciting it to the biexciton state ensures that it will not be excited a second time by the second excitation pulse. This prevents the emission of two pairs, which constitutes an error.}
  \label{fig:excitation}
\end{figure}

For most quantum dots we do not know of really long-lived levels that decay in a cascade, so time-bin entanglement appears to be the only option, which also happens to be more relevant for quantum communication purposes. To achieve time-bin entanglement from a quantum dot the phase difference between the two pump laser pulses has to be carried over to the phase difference between the emitted photon pairs, and thus also intermediately to the phase difference between the two possible excitations of the upper level that will decay in a two-photon cascade. It was originally proposed \cite{Simon05a} that the quantum dot be brought to a metastable state and then further excited to the topmost level of the cascade, the biexciton level as shown in Fig.~\ref{fig:excitation}. Dark excitons were proposed in the same reference as potential metastable states. Because of the difficulty of exciting quantum dots into dark exciton states a simpler version is to go directly from the ground state to the biexciton, with the drawback of a possible second excitation by the late pump pulse. Yet, such an experiment can prove the general possibility of time-bin entanglement.

In either case it is good to keep in mind that in order for the two possible emission cascades to be indistinguishable it is necessary that no trace be left in the pump field or the environment of the quantum dot. This however, does not mean that the biexciton level has to have a coherence time that is long enough to span the gap between the two excitation pulses, but only that its dephasing is not too bad with respect to the emitted wavepacket. The fact that the cascade itself is not always the same, i.e. that the exciton state has a finite lifetime, does not degrade the achievable time-bin entanglement. As in SPDC the phase difference between the exciton and biexciton photon is irrelevant for the entanglement, which depends only on the sum phase. However, as discussed in Ref.~\cite{Simon05a} the uncertainty stemming from the exciton lifetime does lead to an entanglement between the biexciton and exciton photons of a pair. Because this would limit their usefulness for multi-photon protocols such as quantum repeaters, the authors of Ref.~\cite{Simon05a} proposed to employ microcavities to modify the lifetimes such that this unwanted entanglement would be eliminated.

\section{Two-photon coherent excitation of a quantum dot}
\label{sec:twophoton}

The central goal of the photon pair generation from the quantum dot systems is to get exactly one photon at the biexciton and one photon at the exciton frequency that are produced within a short time interval and with a well defined sum phase. This is possible and the exciton and biexciton transition frequencies are well separated due to existence of the biexciton binding energy. Nonetheless, to accomplish the generation process coherently, the quantum dot needs to be excited resonantly.

This task is, despite the favorable energetic structure, not trivial to achieve in epitaxial semiconductor quantum dots. The first, and most important reason is the excess scattered laser light that can easily be much stronger than the single photon signal emitted by the quantum dot. Therefore, the traditional way to excite quantum dots is above-band excitation. Here, one uses a laser with an energy higher than any transition in the quantum dot. This laser creates a multitude of carriers in the vicinity of the quantum dot that can be probabilistically trapped in the quantum dot potential. While it is possible to achieve very high single photon count rates with this method, the probabilistic nature of this process reduces the suitability of such a source for quantum information protocols. Another negative feature of the above-band excitation is related to how exactly the quantum dot levels are populated. Namely, biexciton excitations will be created once the exciton level has been filled and therefore a high rate of biexciton photons requires a very large number of carriers in the quantum dot vicinity. This, however, is very unfavorable because it promotes dephasing of the quantum dot levels due to the electric field fluctuations and causes poor photon statistics properties due to processes like carrier re-capture.

A way to overcome these issues is to exploit the biexciton binding energy, which sets the emission lines of exciton and biexciton photons far apart. When in such a system an excitation laser light is tuned to an energy in between these two energies it produces a resonant two-photon coupling between the ground and the biexciton state. The two-photon approach to excite quantum dots was initially shown \cite{Flissikowski04a} on II-VI quantum dots, but it is quite a bit more difficult to apply it to III-V quantum dots. II-VI quantum dots typically have a much larger biexciton binding energy (the difference between the exciton and the biexciton line can be more than 10~nm), but exhibit otherwise unfavorable optical properties; II-VI quantum dots emit photons in the blue and green spectral range that are, due to losses in the optical fibres, not very suitable for quantum communication. The values for the energy difference between biexciton and exciton lines in III-V quantum dots are in the range of 1-2~nm. Therefore, these systems demand a more thoughtful approach to reduce the laser scattering. Earlier work on III-V quantum dots \cite{Stufler06a} showed the signatures of resonant excitation, like Rabi oscillations, but only in photo-current measurements and not in the optical signal. The first optical measurements under two-photon resonant excitation on III-V quantum dots were shown in Ref.~\cite{Jayakumar13a}. It turns out that this type of excitation also enables and improves several other emission properties compared to the traditional above-band excitation \cite{Predojevic14a,Huber16a} but does not completely remove the blinking due to the random occurrence of charged quantum dot states. This blinking behavior can be improved to some degree by photo-neutralization \cite{Huber16b}.

\begin{figure}
  \centering
  \includegraphics[width=0.9\textwidth]{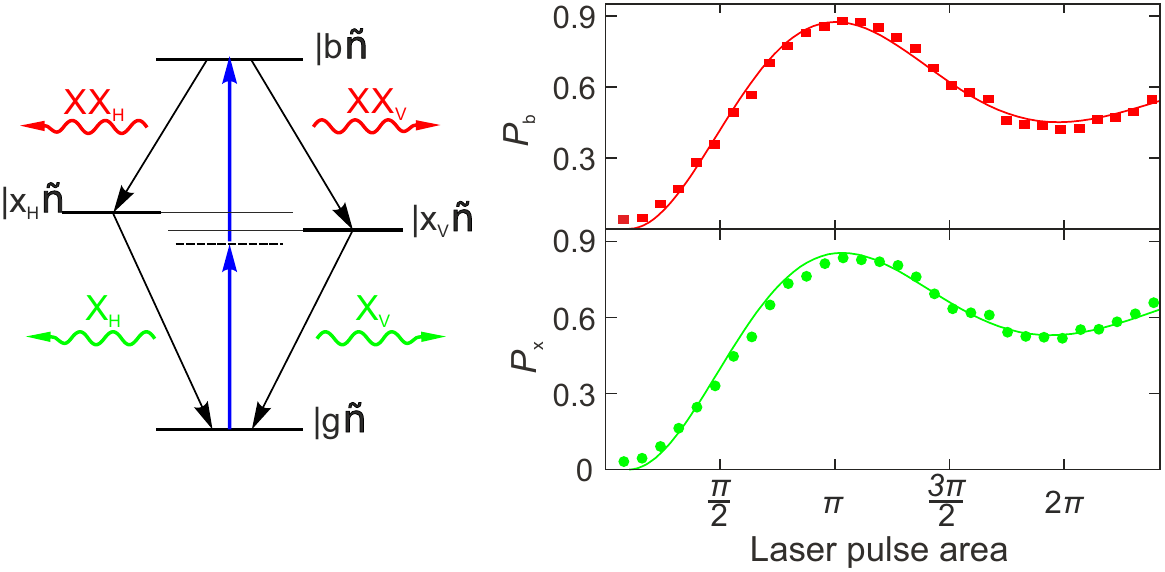}
  \caption{The quantum dot energy scheme on the left shows the exciton fine structure splitting as the energy difference between two exciton levels $\ket{x_{H}}$ and $\ket{x_{V}}$. In the process of two-photon resonant excitation a pulsed laser (shown as arrows pointing upwards) with half the energy of the biexiton state $\ket{b}$ coherently couples the ground ($\ket g$) and biexciton states through a virtual level (dashed line). The biexciton recombination takes place through the intermediate exciton states ($\ket{x_H}$ or $\ket{x_V}$) emitting biexciton ($\mathrm{XX_{H,V}}$) and exciton ($\mathrm{X_{H,V}}$) photons, respectively. On the right the measured biexciton emission probability, $P_b$, and exciton emission probability, $P_x$, as functions of the laser pulse area are compared to a simulation (solid line) that includes linearly intensity-dependent dephasing. The experimental error bars are smaller than the symbols.}
  \label{fig:2p}
\end{figure}

\begin{figure}
  \centering
  \includegraphics[width=\textwidth]{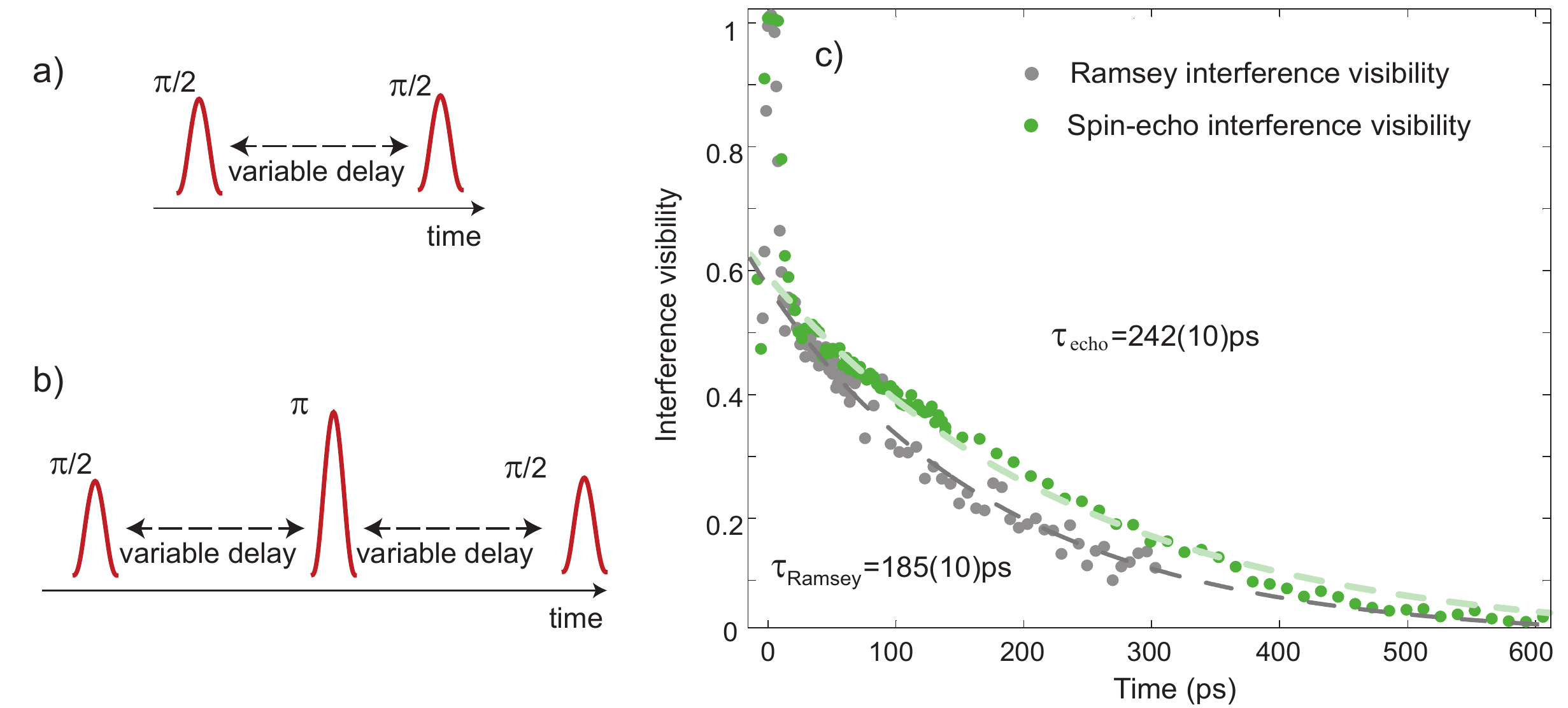}
  \caption{a) Pulse sequence consisting of two $\pi/2$ pulses applied with variable delay. b) The spin-echo pulse sequence. c) The Ramsey interference visibility decay experiment as monitored by the emitted biexciton photons is shown in gray. The data shown in green are from a spin-echo measurement performed on the same emitter.}
  \label{fig:ramsey}
\end{figure}

The coherence of the excitation process enables coherent manipulation of the phase of the ground-biexciton state superposition, which is crucial for obtaining time-bin entanglement as pointed out in the previous section. The traditional way to characterize the coherence between energy levels is to perform a Ramsey interference measurement in which the investigated system is excited using a sequence of two consecutive $\pi/2$ pulses, Fig.~\ref{fig:ramsey}a. The first of these pulses brings the state into a superposition of the ground and biexciton states. Upon this pulse, the system is allowed to evolve freely for a time defined by the variable delay between the pulses, Fig.~\ref{fig:ramsey}. During the free evolution the excitation pseudo-spin precesses along the equator of the Bloch sphere. The second pulse will map the population either back to the ground state or flip it further to the biexciton state, depending on the evolution of the pseudo-spin and the relative phase between the two pulses. A very thorough review of the coherent manipulation of excitons and spins in quantum dot systems is given in Ref.~\cite{Ramsay10a}.

When such an experiment is performed in two-photon excitation it results in Ramsey interference fringes in both the exciton and the biexciton emission \cite{Flissikowski04a}. It is important to note here that in the case of the biexciton emission these fringes are a direct result of the laser driving the transition. The interference observed in the exciton channel closely follows the behavior of the biexciton but comes as a consequence of the cascade decay of the system. The Ramsey interference measurement characterizes the coherence of the ground-biexciton state superposition and by varying the delay between the two Ramsey pulses one can measure the coherence decay of this pseudo-spin. An example of the decay of the Ramsey interference fringe visibility is shown in Fig.~\ref{fig:ramsey}c.

Decoherence caused by low frequency noise can be eliminated by applying a refocusing pulse. Such a measurement is commonly called spin echo (also Hahn echo) and requires a sequence of three consecutive pulses of different intensities ($\pi/2, \pi, \pi/2$) as illustrated in Fig.~\ref{fig:ramsey}b. Due to their lifetime quantum dots are usually excited using laser pulses that are not longer than a few picoseconds. Therefore the simplest way to obtain the sequence of Ramsey pulses is by feeding pulsed laser light into a variable-length Michelson interferometer. Concerning the spin echo measurements, it is quite straightforward to implement such a measurement in systems that have long lifetimes and coherence times. For example, for a trapped ions system where the coherences are of the order of a millisecond one can use light derived from a continuous-wave laser and create the pulse sequence using an acousto-optical modulator. Driving the ground-biexciton state superposition of a quantum dot, however, requires pulse durations of a few picoseconds. In Ref.~\cite{Jayakumar13a} it was shown that the echo sequence with such pulses can be constructed by using a Michelson interferometer in  double-pass configuration. Such an implementation is capable of delivering the three consecutive pulses necessary for the spin-echo sequence with the middle pulse being a result of the interference between the light passing once through the interferometer with the light passing twice. Fig.~\ref{fig:ramsey}c shows two sets of data, one taken in a Ramsey and the other in a spin-echo interference experiment.

The creation of time-bin entanglement requires a phase stable generation of subsequent photon pairs, which can be hampered by the phase uncertainty in the biexciton generation. Therefore, one of the requirements for successful generation of a high degree of time-bin entanglement is the generation of the photon pairs with well defined excitation phase. To predominantly generate single pairs of photons through the biexciton decay, one needs to avoid populating the single exciton state as well as the re-excitation of the biexciton state after a decay within the same laser pulse. This creates conflicting requirements for the excitation pulse length. Namely, short pulses suppress dephasing and decay within the pulse duration, but have large bandwidth and high peak intensity, which increases the off-resonant generation of single excitons and power induced phase shifts. Longer pulses make the system more vulnerable to background dephasing, decay during the pulse and thus multiple excitations.

\begin{figure}
  \centering
  \includegraphics[width=\textwidth]{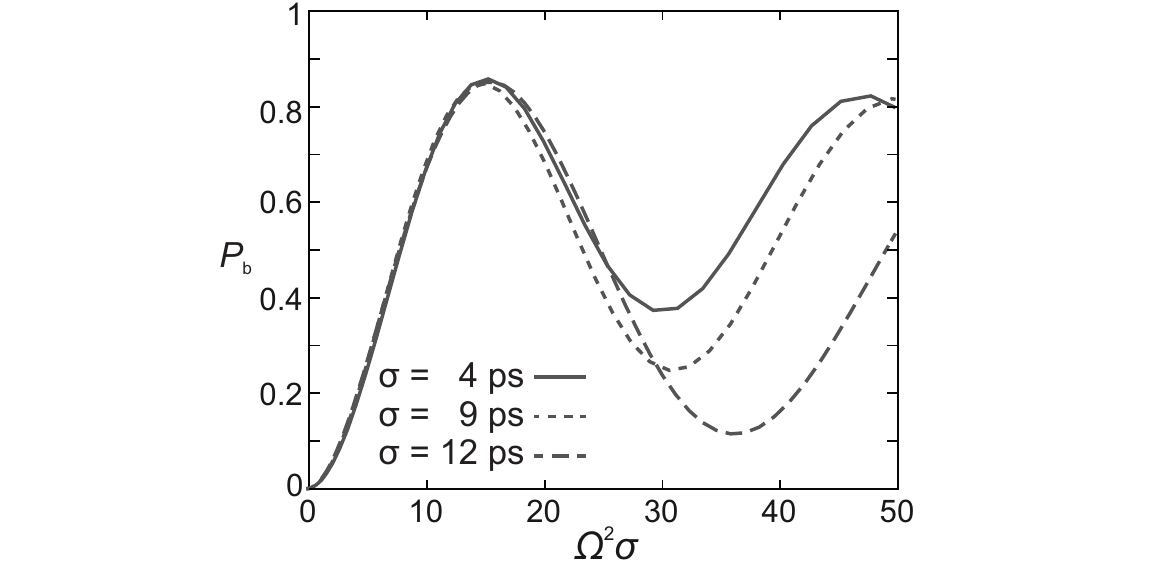}
  \caption{Simulated emission probability for the biexciton, $P_b$ for constant dephasing as a function of $\Omega^2\sigma$, where $\Omega$ is the Rabi frequency and $\sigma$ the pulse length. The damping of the Rabi oscillations strongly depends on the length of the excitation pulse.}
  \label{fig:rabi}
\end{figure}

Nevertheless, one can find an optimized operation regime for the parameters of the system under consideration. The interaction of the quantum dot with the semiconductor environment does not seem to influence this optimization. In Ref.~\cite{Huber16a} we showed that one can choose an excitation pulse length that favors the creation of biexcitons while suppressing the creation of unpaired excitons. This dependence is illustrated in Fig.~\ref{fig:rabi} showing the results of a theoretical simulation conducted in \cite{Huber16a}.

This result has an important consequence. It indicates an existence of a tradeoff between the excitation-pulse length and the biexciton binding energy. In particular it favors the use of quantum dots with large biexciton binding energy that in return allow using short excitation pulses. In addition, such excitation pulses reduce the excitation jitter and are therefore more favorable for quantum information applications.

\section{Time-bin entangled photon pairs from a quantum dot}
Written in terms of the biexciton (XX) and exciton (X) photon modes, the state given in Eq.~(\ref{eq:tbstate}) reads
\begin{equation}
\ket{\Phi}=\frac{1}{2}\left(\ket{E_{XX}E_X}+e^{i\phi_l}\ket{L_{XX}L_X}\right),
\label{eq:psi-time-bin}
\end{equation}
where E(L) denotes the early(late) time bin, XX (X) the biexciton (exciton) recombination photon and $\phi_l$ is the phase between the two pump pulses. In the previous section, we explained how the quantum dot can be excited resonantly. The phase $\phi_l$ in Eq.~(\ref{eq:psi-time-bin}) is the reason why a resonant pumping scheme is necessary. If the pump process is not phase preserving, like above-band excitation, $\phi_l$ will not be the phase between the two pump pulses but some random phase in each emission event, resulting in an overall mixed state. The coherent excitation of the biexciton directly from the ground state, enables the possibility to transfer the phase from the laser onto the quantum dot system  thereby creating an output of the desired form given in Eq.~(\ref{eq:psi-time-bin}).

While still superior to SPDC sources, which emit thermal pair distributions, one drawback of the presented scheme is the inherent creation of four-photon events, even with a perfect quantum dot with zero multi-photon emission. This comes from the fact that the entanglement generation depends on a probabilistic generation of one photon cascade either in the E or in the L time bin. A photon cascade in both of the time bins is a four-photon event, outside the single-pair Hilbert space, and therefore the excitation probability has to be kept at a reasonably low level. The same problem occurs with time-bin entangled photon pairs from SPDC. Contrary to the case of SPDC, two solutions to this problem are known. As discussed above, the first one was proposed by Simon and Poizat~\cite{Simon05a}, which is using a metastable state as the initial state. Thereby, a deterministic creation of the time-bin entangled state is possible, without the loss of other degrees of freedom. This idea has not yet been demonstrated experimentally. The second one, which was already demonstrated experimentally, is to create the entanglement in a different degree of freedom, e.g. polarization, and convert this entanglement to time bin~\cite{Versteegh15a}. This however requires the availability of suitable quantum dots with zero fine-structure splitting, which may have other disadvantages. The conversion  requires that either fast polarization switches are used or an extra 75\% combined loss for the pairs is accepted. A further drawback of this solution is that the simultaneous creation of entanglement in the polarization and time-bin degrees of freedom, so called hyper-entanglement is not possible.

Let us come back to the analysis of the time-bin entanglement, which was already discussed in Section~\ref{sec:time-bin}. As shown in Fig.~\ref{fig:timebinent}, the middle pulses coming out of each analyzing interferometer yield the superposition bases measurements that are important to demonstrate entanglement. As the phases $\phi_1$ and $\phi_2$ are varied, entanglement manifests itself in a variation of the rate of coincidence counts between pairings of two output pulses, one of each interferometer. For a maximally entangled state like Eq.~(\ref{eq:tbstate}) the individual, single count rate would remain constant, independent of the phases, because either photon is individually in a mixed state of the early and late time bins. In coincidence, however, the time bins are interfering, because it is not possible, not even in principle, to tell in which time bin the photon cascade was created and which paths the photons took in the analyzing interferometers. For an imperfect state the coincidence rate will oscillate with an interference visibility that depends on the indistinguishability of the early and late cascades.

\begin{figure}[t]
  \centering
  \includegraphics[width=\textwidth]{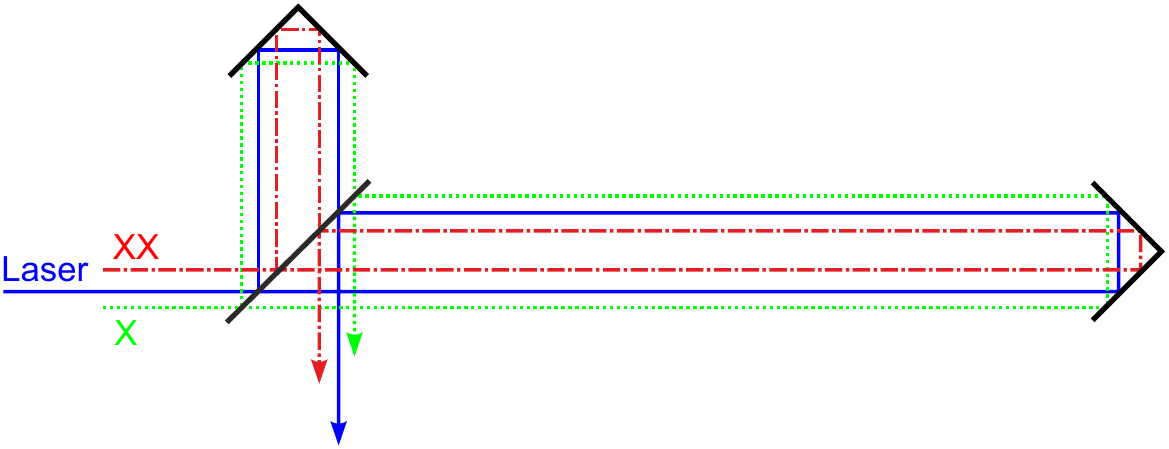}
  \caption{Three different paths in one physical realization of one interferometer were used as the pump interferometer and the three analyzing interferometers.}
  \label{fig:timebininterferometer}
\end{figure}

In our experimental realization a pulsed laser (\SI{80}{MHz} repetition rate, \SI{12}{ps} pulse duration) coherently drove the ground-biexciton transition with a probability of 6\%. To create the two pump pulses, we sent the laser light through an imbalanced Michelson interferometer with a fixed length difference of \SI{1}{m}. This interferometer plus the resonantly pumped quantum dot is the time-bin pair source in Fig.~\ref{fig:timebinent}. After frequency and polarization selection of the XX and X photon, the photons were sent through additional beam paths inside the same physical interferometer (see Fig.~\ref{fig:timebininterferometer}). This ensured the same path length difference for all the three interferometers. Furthermore, any global phase drift would affect all three interferometers equally, thus no stabilization is required. If one wants to use the time-bin entangled photons for any real-world quantum protocol, this interferometer has to be unfolded and stabilized.

\begin{figure}[t]
  \centering
  \includegraphics[width=\textwidth]{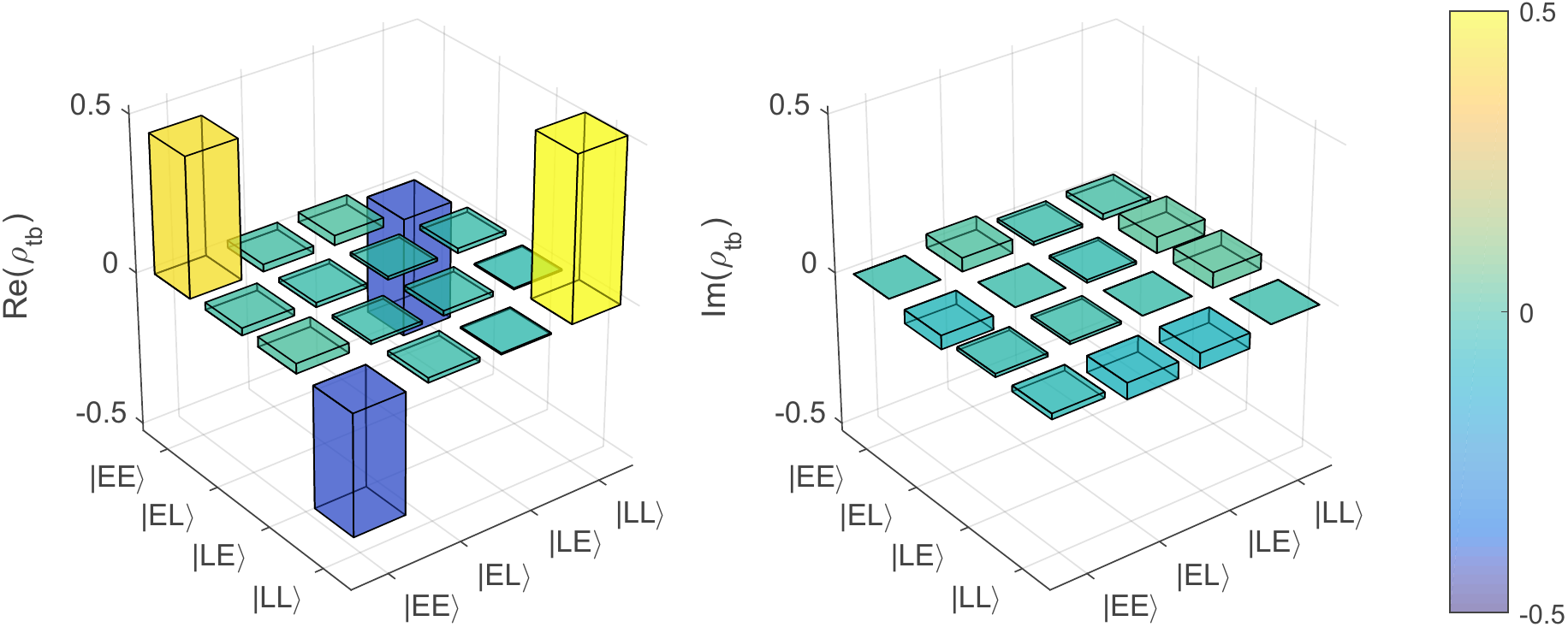}
  \caption{Real and imaginary part of a reconstructed density matrix. This matrix was measured with a \SI{12}{ps} excitation pulse and 6\% excitation probability. \ket{EE} - \ket{LL} denote the measurement basis.}
  \label{fig:densitymatrix}
\end{figure}

For the analysis of the created time-bin entangled state, we used the method of tomographic reconstruction, which needs measurements in a variety of bases, i.e. phase settings. Details on the reconstruction can be found in Ref.~\cite{Jayakumar14a}. The resulting density matrix which is given by $\rho=\ket{\Psi}\bra{\Psi}$ can be seen in Fig.~\ref{fig:densitymatrix}. The diagonal of the matrix from \ket{EE}\bra{EE} to \ket{LL}\bra{LL} represents classical correlations in the E/L basis. The off-diagonal elements are also called the coherences of the state and quantify the entanglement present in the output.

If a source is reasonably close to the desired ideal state it makes sense to quantify the overlap with that state, the so-called fidelity as an elementary measure of the achieved quality. The fidelity $\mathcal{F}$ of an arbitrary mixed state $\rho$ with a pure target state $\ket\psi$ is defined as $\mathcal{F} = \bra\psi \rho \ket\psi$. For the density matrix shown in Fig.~\ref{fig:densitymatrix} the fidelity towards the state $\ket{\Phi^-}$ is $\mathcal{F}=0.88(3)$.

Unfortunately, the quantum dot community has long been using the fidelity as a substitute for a proper entanglement measure. This is not a good practice and it is better to calculate the concurrence, which is defined as
\begin{equation}
\mathcal{C}(\rho)=\mathrm{max}(0,\lambda_1-\lambda_2-\lambda_3-\lambda_4),
\end{equation}
where $\lambda_1,...,\lambda_4$ are the eigenvalues, in decreasing order, of the matrix
\begin{equation}
R=(\sqrt{\rho}\;\widetilde{\rho}\sqrt{\rho})^{1/2},
\end{equation}
where $\widetilde{\rho}=(\sigma_y\otimes\sigma_y)\rho^*(\sigma_y\otimes\sigma_y)$. $\sigma_y$ is the Pauli matrix $\left(\begin{matrix} 0 & -i\\ i& 0\end{matrix}\right)$ for a spin flip and $\rho^*$ is the complex conjugate of $\rho$.
The concurrence is $\mathcal{C}=0$ if no entanglement is present and $\mathcal{C}=1$ for a maximally entangled state. For the density matrix given in Fig.~\ref{fig:densitymatrix} the concurrence is 0.78(6).

These values compare well with earlier achievements in polarization entanglement from quantum dot and allow, at least in principle, a violation of Bell's inequality. For practical applications we would still like to see some improvements. Looking at the density matrix (Fig.~\ref{fig:densitymatrix}) we can identify some shortcomings. First, there is a small imbalance between $\ket{EE}$ and $\ket{LL}$, which is either due to a slightly different pump pulse energy or different transmissivity of the long and short analyzing interferometer arms. Second, we notice that the magnitude of the coherences is smaller than that of the diagonal elements. This is a result of several effects that limit the indistinguishability of the early and late cascade, including dephasing during the excitation process and during the lifetime of the biexciton state. The former effects were discussed in detail in Sec.~\ref{sec:twophoton}. The dephasing during the lifetime of the biexciton is most likely due to the phonon environment remaining at temperatures around \SI{5}{\kelvin} and also due to the fast components (comparable to the biexciton lifetime) of spectral diffusion, which in turn is usually attributed to the fluctuations in the charge environment around the quantum dot. The impact of both these detrimental effects could be reduced most by a lifetime reductions, e.g. using a microcavity and its Purcell effect, but so far no results on resonant two-photon excitation of quantum dots in microcavities have been reported.

\section{Outlook}

The level of time-bin entanglement that has been achieved with quantum dots to date is quite remarkable. Direct single-pair emission, however has not yet been achieved. At this point one needs to work at rather low excitation probabilities and the quantum dot structures that have been used exhibit rather low outcoupling and collection efficiencies. This results in an overall rather low event pair count rate even though the rate of actual excitation events is in the MHz range. This also means that it is difficult to optimize all the relevant parameters for a given quantum dot. For this reason one should apply the same technique to new structures that promise much higher count rates such as nanowire quantum dots \cite{Huber14a} or quantum dot microlenses \cite{Gschrey15a}. The two-photon resonant excitation of these structures may be more difficult, but should be achievable with stronger spectral filtering of the luminescence. Having higher pair count rates will allow investigating the conditions that are required for even better time-bin entanglement. To reduce any unwanted entanglement within a time bin, it would be interesting to also try this out with micropillar microcavities \cite{Schneider09a}. The cavity would be tuned to resonance with the biexciton transition so that the biexciton lifetime is shortened by the Purcell effect as proposed in \cite{Simon05a}. At the same time the increased collection efficiency might make this the perfect time-bin entanglement source.

For creating single time-bin entangled pairs it will be most interesting to investigate the dark exciton preparation \cite{Poem10a} and how to coherently transfer from the dark exciton to the biexciton. Single entangled pairs enable quantum communication protocols with much higher efficiency than entangled pairs from SPDC. Another idea that could readily be demonstrated is the creation of hyperentanglement in the polarization and time-bin degrees of freedom, i.e. a state of the form
\begin{equation}\label{eq:hyper}
  \ket{\psi} = (\ket{HH}+\ket{VV})\otimes(\ket{EE}+\ket{LL}).
\end{equation}
This is useful for certain linear optical quantum information protocols. For example, it is possible to exploit the extra entangled degree of freedom to perform perfect Bell-state analysis, the central process of quantum teleportation and entanglement swapping. Another direction lies in the observation that time-bin encoding and entanglement is not limited to two-dimensional (qubit) configurations. In other systems high-dimensional time encoding has been investigated \cite{Riedmatten02a} but not yet for any single quantum emitter. Finally, if we consider multi-photon entanglement through multilevel cascades in quantum dots or quantum dot molecules, time-bin entanglement may be the only possible way to establish multipartite entangled states such as the GHZ or W states directly from the source.

In summary, the temporal degree of freedom of the photon can be a valuable resource, which has not yet been sufficiently explored for single quantum emitters. It is versatile, because it can apply to any cascaded transition without any particular requirements on energy or spin structure. It does, however require the possibility of coherent control of the topmost energy level of the cascade. In our opinion this is something that ties in with other developments in single emitters, where all properties of single photon sources improve when dedicated coherent interactions are used rather than the primitive above-band pumping. Admittedly, the coherent control increases the complexity of the optical setup, but barring any massive breakthroughs there seems no other way to go.

\bibliographystyle{spphys}
\bibliography{predojevic}

\end{document}